# Code smells

Peter Kokol[1], Milan Zorman[1], Bojan Žlahtič[1], Grega Žlahtič[1]

[1]University of Maribor, Faculty of Electrical Engineering and Computer Science
2000, Maribor, Slovenia
{peter.kokol, milan.zorman, bojan.zlahtic, grega.zlahtic1}@um.si

Fowler [1] defined *code smells* as symptoms of poor design and implementation choices. Such symptoms may originate from activities performed by developers during emergencies, poor design or coding solutions, by making bad decisions, or employing so called anti-patterns [2]. Code smells could also be the consequence of so-called *technical debt* [3]. Among other things, they might hinder comprehension [4] and increase code complexity and fault-proneness and decrease maintainability [5]. To overcome the above problems code smells must be identified and dealt with [6]. Identification of code smells relies on structural information extracted from the source code [7]. Due to the rising number of publications, code smells seem to be a promising approach in software quality assurance. However, there is not yet any descriptive and thematic perspective of the field.

Many modern computer science approaches draw their inspiration from nature [8]. Smells play an important role in communication and assessments of other beings and objects, mainly in mating rituals and searching for food. Since code smells in general carry more or less a negative connotation, we decided to search also for positive code smells that would outline the positive characteristics of the code and, in some cases, like pheromones, attract the customers of software development companies or new developers to open source projects more intensively.

In this manner we decided to perform a bibliometric analysis of the code smell research literature production. Our objective was first to assess the spread of the research, its geographical dispersion and publishing trends, and second to identify main research themes and directions.

## 1. Methods

Bibliometrics has its origins in the beginning of the last century. However, it became »operational« in 1964 with the introduction of the science citation index and prominent because of the need to measure the effects of the large investments going into the research and development. Bibliometrics [9] [10] analyses the properties of literature production in terms of measures, like the number of articles in a scientific discipline, trends of literature production, most prolific or productive entities, most cited papers and authors, etc..

An interesting technique used in bibliometric analysis is bibliometric mapping [11] which visualises literature production based on various text mining techniques [12]. A popular bibliometric mapping software tool is the VOSviewer (Leiden University,

Netherlands) [13]. VOSviewer software extracts, analyses and maps terms or keywords. Different types of maps can be induced.

### 1.1. Data source and corpus

The search was performed on Scopus (Elsevier, Netherlands), the largest abstract and citation database of peer-reviewed literature: scientific journals, books and conference proceedings. The corpus was formed on February 28th, 2017 using the search sting *"code smells"* in information source titles, abstracts, and keywords on all publications covered ny Scopus.

### 1.2. Data extraction and analysis

Using Scopus analysis servies we exported authors affiliations details, source title, publication type and publishing years to Excel (Microsoft, USA) where they were analysed..

Abstracts, titles and keywords were analysed by VOSviewer using default parameters. All common terms like study, baseline, control group, trend, method were excluded from the analysis. Three maps were induced (1) clustered landscapes presenting popularity of terms (more popular terms are presented in larger squares), associations between terms (terms locted near each other are stongly associated) and related term clusters (terms colured with the same colour), (2) timeline landscape presenting the evolution of terms based on the average publishing year of publications in which the term appeared and (3) the auhtors keyword co – ocuurence network presenting which keyords appeared in the same publication (keywords linked together), popularity of a keyord (more popular keywords are presented with larger circles) and citation rate of a keyord based on the average number of citation in which the keyword appeared.

## 2. Results

The search resulted in 337 publications. Among them there were 239 conference papers, 57 articles. 34 conference reviews and 7 other types of publications. More than 70% of publications were published in conference procedings and only about 17% in journals. That might indicate that code smells research is still in the phase of reaching maturity.

First two publication indexed in Scopus were published in 2002 at the conference on reverse engineering. One was the proceding introduction [1] and the other the proceeding paper about detecting code smells during inspections of code written in Java [2]. The first slight rise in research literature productivity was noticed in 2005, the next in 2009 and the last and largest in 2014 (Figure 2.).

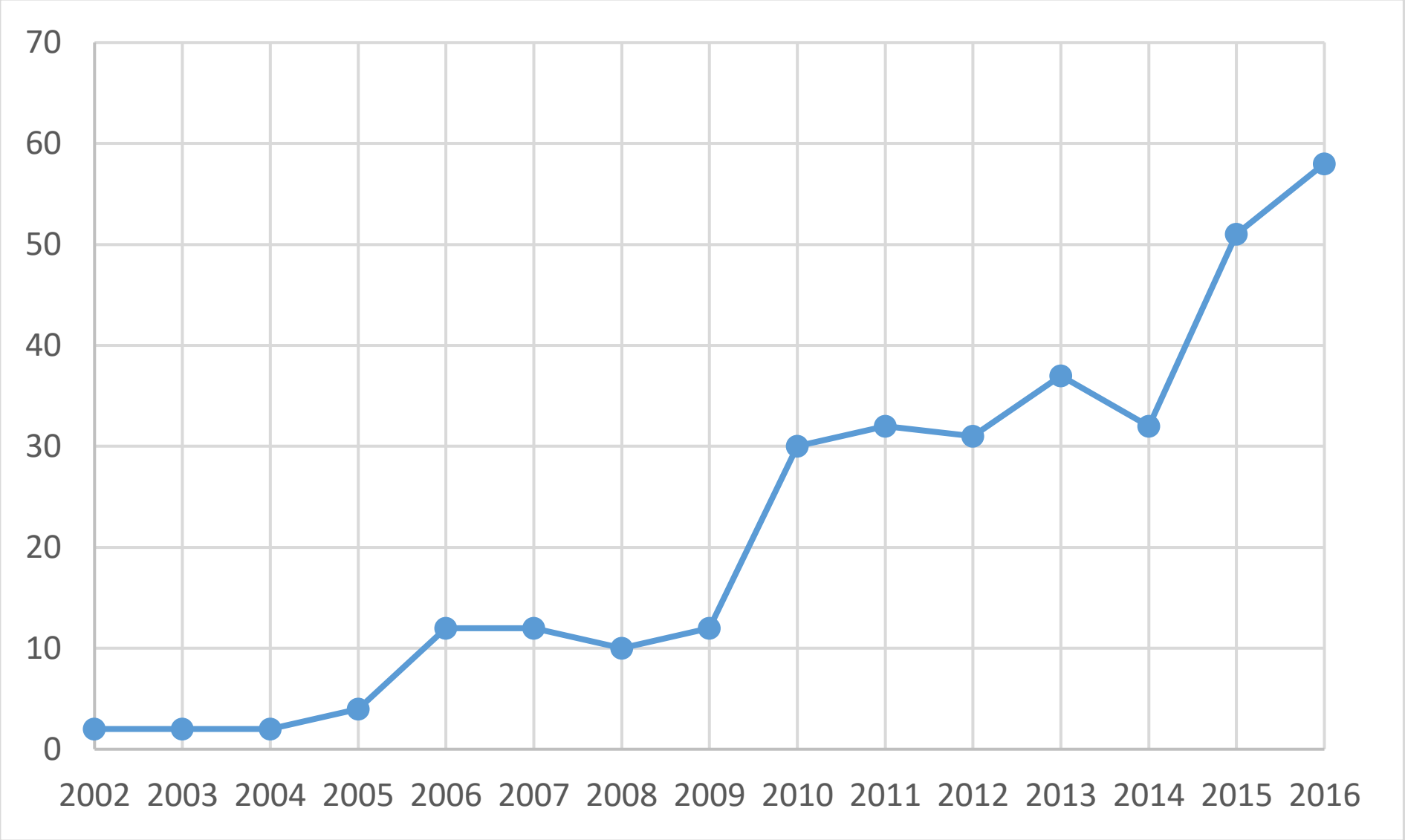

**Fig. 2.** The dynamics of code smells research literature production for the period 2002-2016

Publications appeared in 50 different source titles. The most productive source titles are listed in Table 1.

Table 1. Most productive source titles

| SOURCE TITLE, | NUMBER OF PUBLICATIONS |
|---|---|
| Proceedings International Conference on Software Engineering | 25 |
| Lecture Notes in Computer Science Including Subseries Lecture Notes in Artificial Intelligence and Lecture Notes in Bioinformatics | 23 |
| IEEE International Conference on Software Maintenance Icsm | 13 |
| IEEE Transactions on Software Engineering | 9 |
| Empirical Software Engineering | 6 |
| Journal of Systems and Software | 6 |

The research was distributed over 43 countries and 160 institutions. The 12 most productive countries are shown in Table 2 with the United States being the most productive country. Top 12 countris produced almost 90% of all research literature regarding code smells. Among them there are five G7 countries and also all otherr countries are characterised with strong economies. With the exception of Africa and Australasia all other continents are represented in the list of top productive countries. Henece the research on code smells seems to be globaly widespred, nevertheless the most of the research is performed in most developed countries. The most productive institutions are presented in Table 3. They represent cca. 44% of total research production. While the most productive country is the USA, it is interesting to note, that among the top 14 productive institutions, only one is from the USA and the most (n=3) are from Italy and Europe in general (n=8). That might indicate that the research in code smell is widespred in USA thorough more less productive instutions, contrary to other continents where the research seems to be concentrated in less but more productive institurions. The author keyword network is presented in Figure 3. Five clusters emerged. Based on cluster terms we located and reviewed representative publications, merge related clusters into three and, finally, named them with appropriate research themes (Table 4).

Table 2. Most productive countries

| COUNTRY | NUMBER OF PUBLICATIONS |
|---|---|
| United States | 76 |
| Italy | 39 |
| Brazil | 33 |
| Germany | 30 |
| Canada | 26 |
| India | 18 |
| Norway | 16 |
| Netherlands | 16 |
| United Kingdom | 12 |
| Swiss | 12 |
| China | 11 |
| Japan | 9 |
| TOTAL | 298 |

Table 3. Most productive institutions

| AFFILIATION, | NUMBER OF PUBLICATIONS |
|---|---|
| Universita degli Studi di Milano – Bicocca (Italy) | 19 |
| Pontificia Universidade Catolica do Rio de Janeiro (Brasil) | 15 |
| Universita di Salerno (Italy) | 12 |
| Delft University of Technology (Netherlands) | 12 |
| Universite de Montreal (Canada) | 11 |
| Ecole Polytechnique de Montreal (Canada) | 11 |
| Universita degli Studi del Molise (Italy) | 10 |
| Brunel University London (UK) | 9 |
| Universidade Federal da Bahia (Brasil) | 9 |
| Simula Research Laboratory (Swiss) | 9 |
| Universita degli Studi del Sannio (Italy) | 8 |
| University of Marylandy (USA) | 8 |
| University Oslo (Norway) | 8 |
| Aalto University (Finland) | 8 |
| TOTAL | 149 |

Table 4. Code smells research themes

| Research theme | Scoping review of the theme |
|---|---|
| Code smell detection (yellow cluster) | Code smells are detected using various techniques. Most popular are: Machine learning [14]; detection rules derived from good and bad examples [15], textual analysis [16], metrics [17] and visualisation [18]. The detected smelling code can be used to reduce technical debt [19] |
| Bad smell based software refactoring (violet and blue colours) | Refactoring is one key issue to increase internal software quality and maintainability. Bad smells are used to identify structures where refactorings should be applied [20] [21]. Automatic tolls can be used [22] [23]. Examples of bad smells are duplicated and cloned code, code fragments and similar [24] [25]. |
| Software development and anti-patterns (red and green colours) | The concept of anti – patterns and code smells can utilise the knowledge of known software development problems and improve the quality of products [26] |

As the identified main research themes show, there is a growing concern about code

»quality« and that is for a good reason. In modern computing [27] , where the amount of data to be processed grows by the hour and hardware capabilities are almost at a standstill, there is a growing need for optimization and improvement of code. Code small detection, Bad smell based software refactoring, Software development and anti-patterns are concepts that can be used to improve software performance, usability and maintainability.

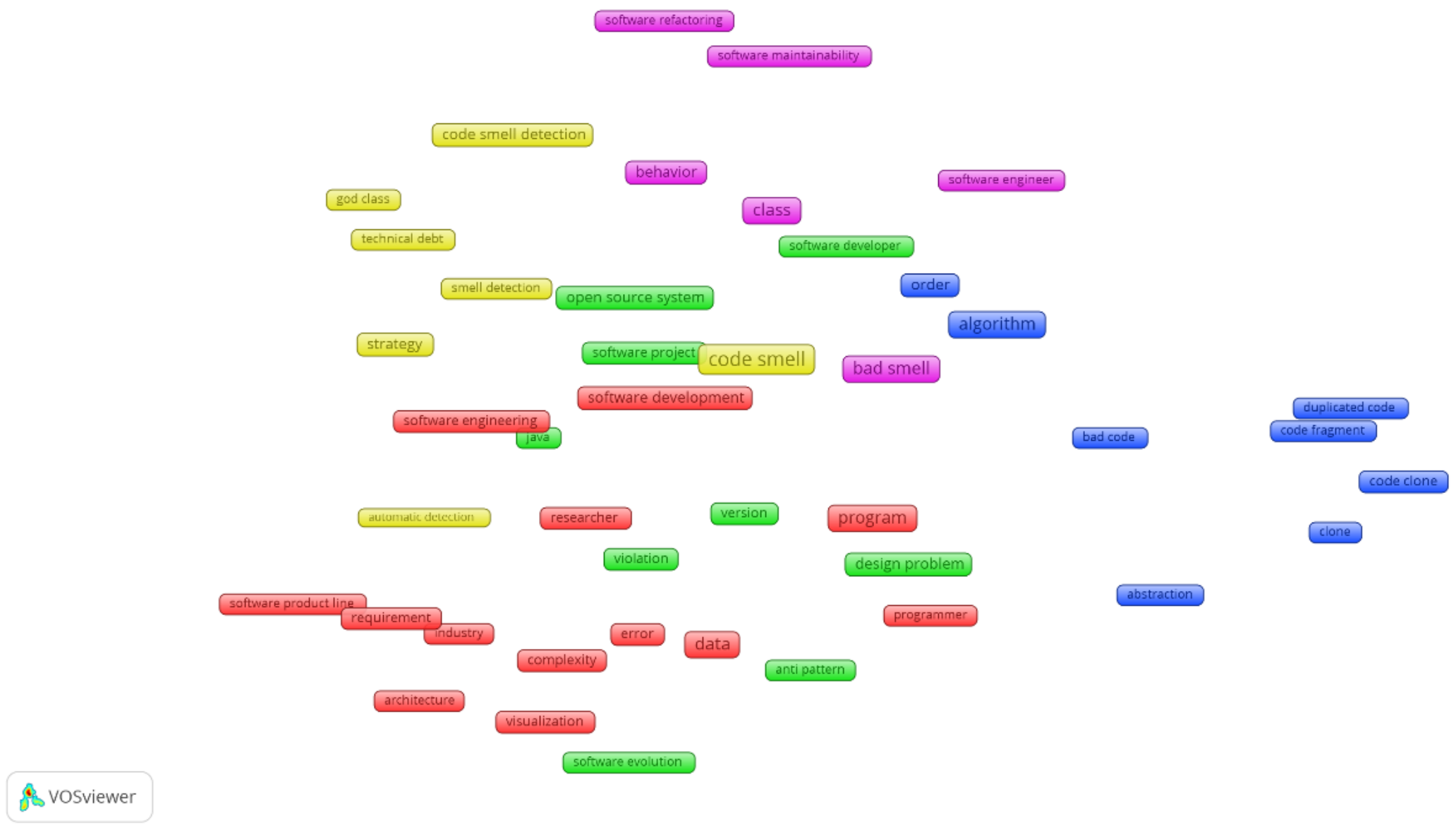

**Fig. 3**. Authors keyword network

## 3. Conclusions and future directions

Our study showed that the interest in code smells research is increasing. However, most of the publications are appearing in conference proceedings, a fact which might reflect that the research has not yet reached the mature phase. The most of the research is done in G7 and other highly developed countries. It seems that the research in the USA is more dispersed than in other highly productive countries, which have one or two strong research centres dealing with code smell investigations. The results show that code smells can also have a positive connotation – we can develop software which “smells good” and attracts various customers and good smelling code could also serve as a pattern for future software development. We also identified some gaps which, in a positive manner, can serve as future research directions. The research is focused mainly on object oriented code, while many software systems are still coded in procedural and other paradigms. Identifying “good smells“ could be another area worth researching. Universal metrics independent of the type of the source text (specifications, architecture, requirements, code, visual presentations, etc) and programming paradigms could be used to identify other types of smells like the requirement smell, functionality smells, user interface smells, and similar [28].

## 4. References

[1] M. Fowler, Refactoring: Improving the Design of Existing Code., Reading: Addison - Wesley, 1999.

[2] V. Arnaoudova, M. Di Penta and G. Antoniol, “Linguistic antipatterns: what they are and how developers perceive them,” *Empirical Software Engineering,* vol. 21, no. 1, pp. 104 - 158, 2016.

[3] F. Shull, „Perfectionists in a World,“ *IEEE Software,* Izv. 28, št. 2, pp. 5 - 6, 2011.

[4] B. Walter and T. Alkheair, “The relationship between design patterns and code smells: An exploratory study,” *Information and Software Technology,* vol. 74, pp. 127 - 142, 2016.

[5] W. Fenske and S. Schulztze, “Code Smells Revisited: A Variability Perspective,” in *Proceedings of the Ninth International Workshop on Variability Modelling of Software-intensive Systems* , Hildesheim, 2015.

[6] F. A. Fontana, M. Mangiacavalli, D. Pochiero in M. Zanoni, „On experimenting refactoring tools to remove code smells,“ v *Scientific Workshop Proceedings of the XP2015* , Helsinki, 2015.

[7] F. Palomba, M. Di Penta, R. Oliveto in D. Poshyvanyk, „Mining Version Histories for Detecting Code Smells,“ *IEEE TRANSACTIONS ON SOFTWARE ENGINEERING,* Izv. 41, št. 5, pp. 462 - 489, 2015.

[8] N. Siddique and H. Adeli, "Nature Inspired Computing: An Overview and Some Future Directions," *Cognitive Computation,* vol. 7, no. 6, pp. 706 - 714, 2015.

[9] E. Garfield, "The History and Meaning of the Journal Impact Factor," *Journal of American Medical Association,* vol. 295, no. 1, pp. 90-93, 2006.

[10] N. De Bellis, Bibliometrics and Citation Analysis, Lanham, Maryland, Toronto, Plymouth, UK: The Scarecrow Press, Inc., 2009.

[11] "Liu, D D; Liu, S L; Zhang, J H," *Medical Devices: Evidence and Research,* vol. 7, pp. 357 - 361, 2014.

[12] M. Sedighi and A. Jalalimanesh, "Mapping research trends in the field of knowledge management".

[13] N. J. vanEck and L. Waltman, "Software survey: VOSviewer, a computer program for bibliometric mapping," *Scientometrics,* vol. 84, no. 2, pp. 523 - 538, 2010.

[14] A. Fontana, M. V. Mantyla, M. Zanoni in A. Mariano, „Comparing and experimenting machine learning techniques for code smell detection,“ *Empirical Software Engineering,* Izv. 21, št. 3, pp. 1143 - 1191, 2016.

[15] U. Mansoor, M. Kassentini, B. R. Maxim in K. Deb, „Multi-objective code-smells detection using good and bad design examples,“ *Software Quality Journal,* pp. 1 - 24, 2016.

[16] F. Palomba, „Textual Analysis for Code Smell Detection,“ v *International Conference on Software Engineering*, Florence, 2015.

[17] B. Walter, B. Matuszyk in F. A. Fontana, „Including structural factors into the metrics-based code smells detection," v *16th International Conference on Agile Software Development*, Heslinki, 2015.

[18] S. J. Lee, X. Lin, L. H. Lo, Y. C. Chen in J. Lee, „Code smells detection and visualization of software systems," v *International Computer Symposium*, Taichung, Taiwan, 2014.

[19] F. A. Fontana, V. Ferme in S. Spinelli, „Investigating the impact of code smells debt on quality code evaluation," v *nternational Workshop on Managing Technical Debt*, Zurich; Switzerland, 2012.

[20] M. W. Mkaouer, M. C. M. ¸. S. Kassentini in K. Deb, „A robust multi-objective approach to balance severity and importance of refactoring opportunities," *Empirical Software Engineering,* pp. 1-34, 2016.

[21] F. Simon, F. Steinbrückner in C. Lewerentz, „Metrics based refactoring," v *5th European Conference on Software Maintenance and Reengineering*, Lisbon, 2001.

[22] g. Szoke, C. Nagy, L. J. Fulop, F. R in T. Gyimóthy, „FaultBuster: An automatic code smell refactoring toolset," v *International Working Conference on Source Code Analysis and Manipulation*, Bremen, Germany, 2015.

[23] S. Vidal, H. Vazquez, J. A. Diaz-Pace, C. Marcos, A. Garcia and W. Oizumi, "JSpIRIT: A flexible tool for the analysis of code smells," in *International Conference of the Chilean Computer Science Society*, Santiago, Chilie, 2015.

[24] F. Meng, „dentifying refactoring opportunities from code clones based on SOM clustering," *ICIC Express Letters, Part B: Applications,* Izv. 5, št. 2, pp. 1087 - 1092, 2014.

[25] A. Ouni, M. Kessentini, S. Bechikh and H. Sahraoui, “Prioritizing code - smells correction task useing chemical reaction optimization,” *Software Quality Journal,* vol. 23, no. 2, pp. 323 - 361, 2015.

[26] Y. Luo, A. Hoss in D. L. Carver, „An ontological identification of relationships between anti-patterns and code smells,“ v *IEEE Aerospace Conference*, Big SKy, United States, 2010.

[27] J. S. Saltz, „The need for new processes, methodologies and tools to support big data teams and improve big data project effectiveness,“ v *Big Data (Big Data), 2015 IEEE International Conference on*, 2015.

[28] M. Pighin, V. Podgorelec and P. Kokol, “Fault-Threshold Prediction with Linear Programming,” *Empirical Sofware Engineering,* vol. 8, no. 2, pp. 117 - 138, 2003.